\documentstyle[amssymb,preprint,12pt,aps,epsf]{revtex}
\begin{document}
%
%\euro{}{}{}{}
%\Date{}
%\shorttitle{A. Pimenov {\it et al.} Universal Relationship between etc.}
%
\title{Universal relationship between the penetration depth
 and the normal-state conductivity in YBaCuO.}
\author{A. Pimenov$^{1}$, A. Loidl$^{1}$, B. Schey$^{1}$, B. Stritzker$^{1}$, G.
Jakob$^{2}$, H. Adrian$^{2}$,
A. V. Pronin$^{3}$, and Yu. G. Goncharov$^{3}$}
\address {$^{1}$Institut f\"{u}r Physik, Universit\"{a}t Augsburg, 86135
Augsburg, Germany\\
$^{2}$Institut f\"{u}r Physik, Universit\"{a}t Mainz, 55099 Mainz, Germany%
\\
$^{3}$Institute of General Physics, Russian Acad. Sci., 117942 Moscow, Russia%
}
%\rec{}{}
%\pacs{
%\Pacs{74}{25.-q}{General Properties}
%\Pacs{74}{25.Fy}{Transport Properties}
%\Pacs{74}{76.Bz}{High-Tc films}
%      }

\maketitle

\begin{abstract}

The absolute values of the conductivity in the normal state $\sigma _{n}$
and of the low temperature penetration depths $\lambda (0)$ were measured
for a number of different samples of the YBaCuO family. We found a striking
correlation between $\sigma _{n}$ and $\lambda ^{-2}\,$ regardless of
doping, oxygen reduction or defects, thus providing a simple method to
predict the superconducting penetration depth and to have an estimate of the
sample quality by measuring the normal-state conductivity.

\end{abstract}

The superconducting penetration depth, $\lambda (0),$ is one of the most
important electrodynamic properties of the high-temperature superconductors
and provides direct information about the spectral weight of the
superconducting condensate \cite{microwave}. In the normal state the low
frequency electrodynamics can be characterized by the absolute value of the
conductivity $\sigma _{n}(\nu ,T)$. At not too high frequencies,
$\sigma _{n}(\nu ,T)$ is
in a good approximation a real function of
frequency. Because $\lambda (0)$ and $\sigma _{n}(0,T)$ both involve a
Fermi-surface averaging and depend on the same amplitude factors $%
e^{2}N(0)\upsilon _{{\rm {F}}}^{2}$ \cite{hensen}, they may be expected to
be connected to each other.

In order to investigate possible correlations experimentally, we have
carried out submillimeter-wave (100 GHz < $\nu $ <%
1100 GHz) transmission experiments on different thin YBa$_{2}$Cu$_{3}$O$%
_{7-\delta }$ films. Optimally doped films on different substrates were
obtained by laser ablation and by magnetron sputtering. One film was oxygen
reduced after the measurement and then remeasured under identical
conditions. Transmission experiments in the frequency range 100-1100 GHz
were performed utilizing a set of backward-wave oscillators. The
measurements were performed in a Mach-Zehnder interferometer arrangement
\cite{volkov} which allows both, the measurements of transmission and phase
shift. The properties of the substrates were determined in separate
experiments. Utilizing the Fresnel optical formulas for the complex
transmission coefficient for a double layer system, the complex conductivity
was determined from the observed spectra without any approximations. It is
important to note, that the absolute values of the conductivity could be
deduced directly from the experiment. The penetration depth was calculated
from the conductivity data via $\lambda =c/\omega k$, with $k$ being the
imaginary part of the complex refractive index $k=Im[i(\sigma
_{1}+i\sigma _{2})/\varepsilon _{0}\omega ]^{1/2}$ . This expression for $%
\lambda $ gives the usual microwave result $\lambda =(\mu _{0}\omega \sigma
_{2})^{-1/2}$ in the limit $\sigma _{1}\ll \left| \sigma _{2}\right| $ and
reduces to the expression for the skin depth in the normal state ($\sigma
_{1}\gg \left| \sigma _{2}\right| )$.

The submillimeter properties of YBa$_{2}$Cu$_{3}$O$_{7-\delta }$ films for a
fixed frequency $\nu =450$ $GHz$ (15 cm$^{-1}$) are summarized in the Table
1. In addition, some details of the film preparation, the
weight of the superconducting condensate ($\omega _{p,s}^{2}=c^{2}/\lambda
(0)^{2}$), and the effective scattering rate (Eq. 1) are given.
The detailed submillimeter-wave data for the
samples No.1,3,5 were published previously \cite{recent,biegel,reduced}. The
data in Tab.1 are represented in the order of decreasing absolute values of
the conductivities at 100 K. Comparing the conductivity column with the
penetration depth column, the correlation between these two
quantities becomes obvious. With the exception of the sample No.4, the
penetration depth increases with decreasing conductivities. We
note that the scattering of data for different samples may be
connected with experimental difficulties in measuring the absolute
values of conductivity and penetration depth. In addition, in  spite of
substantial progress in prepairing good quality YBaCuO films
\cite{films}, the details of deposition process still may influence
the experimental results.

To check the possible correlation between $\sigma _{n}$ and $\lambda $ in
more detail, the normal state conductivity and the superconducting
penetration depth were analyzed on the basis of published data available up
to now. The results of this analysis are summarized in Fig. 1. In order to
simplify the analysis only results on YBaCuO based materials were
considered. To compare the absolute values of the conductivity for different
samples the characteristic temperature of $T$=100 K was chosen. It is well
known \cite{iye} that the resistivity even of oxygen reduced or doped
cuprates is to a reasonable approximation linear for temperatures between
100 K and 300 K. A linear temperature scaling was therefore used, if the
results were reported at temperatures different from $T$=100 K. Taking into
account the relatively strong scattering of the data for different samples,
this procedure certainly introduces no substantial errors.

Fig. 1 represents the data as measured by different experimental methods. In
addition to the thin film results by THz transmission technique \cite
{brorson,feenstra,ludwig,nagashima,dahne,nuss,vaulchier}, mutual inductance
data \cite{fiory,ulm} and single crystal data \cite
{basov,zhang,kitano,sonier,kamal} are shown. The remarkable feature of Fig. 1 is
that the majority of experimental points closely follows the dashed line,
which is the best fit to our data according to the expression $\lambda
(0)^{-2}/\mu _{0}=\sigma _{n}/\tau $, with 1/$\tau =22$ THz. Fig. 1 thus
suggest that the absolute value of the conductivity in the normal state is
approximately proportional to the spectral weight of the superconducting
condensate $\omega _{p,s}^{2}$ or to the inverse square of the penetration
depth $\lambda (0)^{-2}$.

In the following we would like to suggest rather
simple arguments to understand this relation.
Within the isotropic approximation the real part of the
conductivity can be written as $\sigma _{n}=\varepsilon _{0}\omega
_{p,n}^{2}\tau $, where $\tau ^{-1}$ is the quasiparticle scattering rate
and $\varepsilon _{0}\omega _{p,n}^{2}$ is the spectral weight of the Drude
peak. For a number of high temperature cuprates in the normal conducting
state the characteristic scattering rate has been shown to increase linear
in temperature following $%TCIMACRO{\UNICODE[m]{0x127}}
%BeginExpansion
\rlap{\protect\rule[1.1ex]{.325em}{.1ex}}h%
%EndExpansion
/\tau \approx 2k_{{\rm {B}}}T$ \cite{ir}. Assuming the value $\Delta =2k_{%
{\rm {B}}}T_{{\rm {C}}}$ for the energy gap \cite{gap}, one obtains the
approximation for the ratio of mean free path to the coherence length, $\ell
/\xi =\pi \tau \Delta /\hslash \approx 3$. As the scattering rate decreases
even stronger than linearly below $T_{{\rm {C}}}$ \cite{microwave}, at low
temperatures the high-$T_{{\rm {C}}}$ superconductors are in the clean limit
and the spectral weight of a Drude peak $(\varepsilon _{0}\omega
_{p,n}^{2}=ne^{2}/m^{*})$  is expected to fully
condense into the delta function at $T$=0K. This conclusion is supported by
the Drude-analysis of the frequency dependent conductivity of YBa$_{2}$Cu$%
_{3}$O$_{7-\delta }$ \cite{recent} and by a comparison of the conductivity
spectral weights at infrared frequencies \cite{tanner}. In the clean limit
the following expression holds

\begin{equation}
\frac{1}{\mu _{0}\lambda (0)^{2}}=\varepsilon _{0}\omega
_{p,s}^{2}=\varepsilon _{0}\omega _{p,n}^{2}=\sigma _{n}/\tau
\end{equation}

The Eq.(1) thus suggests the possible explanation of the correlation in Fig. 1:
obviously defects (doping, sample quality) and oxygen concentration change
only the density of states at the Fermi-level, N(0), or the
effective number of charge carriers, $\omega _{p,s}^{2}$ and $%
\omega _{p,n}^{2}$, and do not affect the effective scattering rate $\tau
^{-1}$. Theoretical calculations of the normal-state properties
using spin-fluctuation scattering \cite{hirschfeld}, Fermi-surface nesting
\cite{nested} or
phenomenological marginal Fermi-liquid approach \cite{varma} result in the
quasiparticle scattering rate, which is not dependent on N(0), but is
approximately proportional to temperature. Therefore, within a
reasonable approximation, $\tau^{-1}$ may be supposed to be  unsensitve
to the level of defects.
As long as the sample is in the clean limit, the complete Drude
spectral weight $\omega _{p,n}^{2}$ condenses into the superconducting delta
function retaining the correlation between the normal and the
superconducting state given by Eq. (1).

Interestingly, the Ni- and Zn-doped samples of Ulm et al. \cite{ulm} and the
Pr doped samples of Brorson et al. \cite{brorson} still follow the
correlation of Fig.1, except for the samples with the highest doping levels
(6\% Ni, 40\% Pr, or 4\% Zn doping). The different behavior of heavily doped
samples is probably due to the fact, that they are no more in the clean
limit. Therefore a substantial portion of the ''normal'' spectral weight is
not condensed into the delta function at low temperatures and the
penetration depth $\lambda (0)=c/\omega _{p,s}$ is substantially higher than
the value estimated from Eq. (1), $\lambda (0)=c/\omega _{p,n}$. In addition
Fig. 1 shows that Zn doping has the strongest effect on the electrodynamic
properties of high-T$_{{\rm {C}}}$'s compared to Ni or Pr. Similar
conclusion has been drawn on the basis of the microwave data by Bonn et al.
\cite{bonn}.

Although simple arguments, presented above, give an idea to
understand the observed correlation, we cannot exclude other
possible explanations. E.g. even the application of Fermi-liquid
picture to high-temperature superconductors is still an
intensively
discussed problem \cite{hensen,shulga}. Further on and
especially concerning  optimally doped  YBCO
films, the extrinsic effects may also play an important role in
determining the conductivity values. The influence of weak links
on the surface impedance of granular superconductors was discussed in detail by
Halbritter \cite{halbritter}.

The scattering rate 1/$\tau =22$ THz, as deduced from the dashed line in
Fig. 1, leads to the following expression at $T$=100 K: $
%TCIMACRO{\UNICODE[m]{0x127}}
%BeginExpansion
\rlap{\protect\rule[1.1ex]{.325em}{.1ex}}h%
%EndExpansion
/\tau =1.7k_{{\rm {B}}}T$. This relation is similar to $
%TCIMACRO{\UNICODE[m]{0x127}}
%BeginExpansion
\rlap{\protect\rule[1.1ex]{.325em}{.1ex}}h%
%EndExpansion
/\tau \approx 2k_{{\rm {B}}}T$ \cite{ir} cited above. On the basis of the
above presented discussion we may conclude, that the scattering rate $
%TCIMACRO{\UNICODE[m]{0x127}}
%BeginExpansion
\rlap{\protect\rule[1.1ex]{.325em}{.1ex}}h%
%EndExpansion
/\tau =1.7k_{{\rm {B}}}T$ is to a good approximation independent of doping,
oxygen depletion and defects. Hence, the proportionality
between $\lambda (0)^{-2}$ and $\sigma _{n}$
provides a simple possibility to characterize the superconducting properties
of the YBaCuO samples by measuring the normal state conductivity by e.g.
standard four point method. On the basis of data, presented in Fig.1 the
penetration depth of a certain sample is given by the approximate expression

\begin{equation}
\lambda (0)[nm]=190/\sqrt{\sigma _{n}(100K)[10^{4}\Omega ^{-1}cm^{-1}]}%
=19\cdot \sqrt{\rho _{n}(100K)[\mu \Omega \cdot cm]}
\end{equation}

Eq. (2) describes 80\% of all penetration depth data in Fig. 1 with
deviations well below 25\%. Finally it should be noted, that most points of
Fig.1 represent thin-film data. The applicability of Eq. (2) for single
crystals should be checked in more detail in further investigations.

In conclusion, we have measured the conductivity and the penetration depth
of a variety of YBaCuO samples and compared our results to published data.
We have plotted the low temperature penetration depth as a function of the
normal-state conductivity, where the absolute values for both parameters
were available on the same sample. In this plot we observe a correlation $%
\lambda ^{-2}$ $\backsim \sigma _{n}$. This observation allows the estimate
of the penetration depth of a given YBaCuO sample by measuring its normal
state conductivity. In the normal conducting state we find universal
scattering rate $%TCIMACRO{\UNICODE[m]{0x127}}
%BeginExpansion
\rlap{\protect\rule[1.1ex]{.325em}{.1ex}}h%
%EndExpansion
/\tau =1.7k_{{\rm {B}}}T$ independent of defects, oxygen concentration and
doping.

This research was partly supported by the BMBF under the contract number
13N6917/0 (EKM).

\newpage
\begin{tabular}{|c||c|c|c|c|c|c|c|c|c|}
\hline
No. & $T_{\text{C}}$ [K] & Preparation & Substrate & $
\begin{array}{c}
\text{d} \\
\lbrack \text{nm}]
\end{array}
$ & $
\begin{array}{c}
\sigma _{n} \\
\text{\lbrack 10}^{4}\Omega ^{-1}\text{cm}^{-1}\text{]}
\end{array}
$ & $
\begin{array}{c}
\lambda  \\
\lbrack \text{nm}]
\end{array}
$ & $
\begin{array}{c}
\omega _{p,s}^{2} \\
\lbrack e{\rm V^{2}}]
\end{array}
$ & $
\begin{array}{c}
\tau ^{\text{-1}} \\
\lbrack \text{THz}]
\end{array}
$ & Ref. \\ \hline\hline
1 & 89.5 & magnetron sputt. & NdGaO$_{3}$ & 81 & 1.51 & 152 & 1.71 & 22.8 &
[4] \\ \hline
2 & 91.3 & magnetron sputt. & LaAlO$_{3}$ & 90 & 1.0 & 204 & 0.952 & 19.1 &
\\ \hline
3 & 85.9 & laser ablation & MgO & 85 & 0.80 & 243 & 0.671 & 16.8 & [5] \\
\hline
4 & 89 & magnetron sputt. & NdGaO$_{3}$ & 75 & 0.64 & 215 & 0.857 & 26.9 &
\\ \hline
5 & 56.5 & No.1 reduced & NdGaO$_{3}$ & 81 & 0.60 & 279 & 0.509 & 17.0 & [6]
\\ \hline
6 & 91 & laser ablation & NdGaO$_{3}$ & 70 & 0.38 & 292 & 0.464 & 24.6 &  \\
\hline
\end{tabular}

\bigskip

\bigskip

\bigskip

\bigskip

\bigskip
Table 1. Submillimeter-wave properties of YBa$_{2}$Cu$_{3}$O$_{7-\delta }$
films at $\nu =450$ GHz. The normal state conductivity $\sigma _{n}$ and the
superconducting penetration depth $\lambda $ are given at temperatures 100 K
and 6 K respectively. The spectral weight of the superconducting condensate $%
\omega _{p,s}^{2}$ and the effective scattering rate 1/$\tau$ were calculated using Eq. (1).

\newpage

Figure caption.

Fig. 1. Low temperature ($T\lesssim 10$ K) penetration depth $\lambda $ of
different YBCO samples as a function of the normal-state conductivity $%
\sigma _{n}$ at $T$=100 K. Dashed line is drawn according to the expression $%
\lambda (0)^{-2}/\mu _{0}=\sigma _{n}/\tau $ using 1/$\tau =22$THz. The
characters in the symbols correspond to:

Rhombs - single crystal data:

B - Basov et al. \cite{basov} and Zhang et al. \cite{zhang}; K - Kitano et
al. \cite{kitano}; K1 - Kamal et al. \cite{kamal}; S - Sonier et al. \cite{sonier}.

Circles - optimally doped YBa$_{2}$Cu$_{3}$O$_{7-\delta }$ films:

closed circles - this work, Table 1; B - Brorson et al. \cite{brorson}; D -
D\"{a}hne et al. \cite{dahne}; F - Fiory et al. \cite{fiory}; L - Ludwig et
al. \cite{ludwig}; N - Nagashima et al. \cite{nagashima} ($\lambda $ is
taken at $T$=40 K); N1 - Nuss et al. \cite{nuss}; V - de Vaulchier et al.
\cite{vaulchier}.

Squares - oxygen reduced YBa$_{2}$Cu$_{3}$O$_{7-\delta }$ films:

closed square - this work, Table 1; L - Ludwig et al. \cite{ludwig}.

Down triangles - doped YBa$_{2}$Cu$_{3}$O$_{7-\delta }$ films:

B - Brorson et al. \cite{brorson} (20\%, 30\%, and 40\% Pr - doped films); F
- Feenstra et al. \cite{feenstra} (DyBa$_{2}$Cu$_{3}$O$_{7-\delta }$); U -
Ulm et al. \cite{ulm} (2\%, 4\% and 6\% Ni doped and 2\% and 4\% Zn doped
films). Dotted lines are guide to the eye. Arrows indicate the increasing
doping direction.

\end{document}